\documentclass[11pt]{amsart}

\usepackage{times,amssymb,amsmath,float,units}
\usepackage{euscript} \usepackage[dvips]{epsfig}

\newcommand{\binf}{\mathbb{F}_2} 
 
\newcommand{\ncr}[2]{\left (\negthickspace \begin{array}{c} #1 \\ #2
\end{array} \negthickspace \right )}
\newcommand{\expec}[1]{\mathbb{E} \left ( #1 \right ) }
 \newcommand{\prob}[1]{\mathbb{P} \left
( #1 \right ) } \newcommand{\probi}[1]{\mathbb{P}_i \left ( #1 \right
) } 

\newcommand{\ent}[1]{h\! \left (\! #1 \! \right)}
\newtheorem{proposition}{Proposition}
\newtheorem{theorem}[proposition]{Theorem}

\newtheorem{lemma}[proposition]{Lemma}

\newcommand{\remove}[1]{} \newcommand\nc\newcommand

 \def\dist{\qopname\relax{no}{d}}

\newcommand\reals{{\mathbb R}} 
 
 \nc\dgv{\delta_{\text{\rm GV}}}
\nc\dlp{\delta_{\text{\rm LP}}} \nc\rcrit{R_{\text{crit}}}
\renewcommand\epsilon{\varepsilon}

\newcommand{\beeq}{\begin{eqnarray*}}
\newcommand{\eneq}{\end{eqnarray*}}

\newcommand\half{\nicefrac{1}{2}}

\begin{document}
\title[Distance distribution of codes and error probability] {Distance
distribution of binary codes and the error probability of decoding}
\thanks{The results of this paper were presented in part at the
International Workshop on Coding and Cryptography, March 2003, Paris,
France and at the 2003 IEEE International Symposium on Information
Theory, June 29 - July 4, 2003, Yokohama, Japan. } 

\author[Alexander Barg]{Alexander Barg$^{\ast}$} 
\thanks{$^{\ast}$ 
Dept. of ECE, University of Maryland,  College Park, MD 20742. E-mail
{\tt abarg@umd.edu}.  Research partly done at DIMACS, Rutgers
University, 96 Frelinghuysen Rd.,  Piscataway, NJ 08854. Research
supported in part by NSF Grant CCR-0310961 and by Minta Martin 
Aeronautical Research Fund of 
the University of Maryland.}  

\author[Andrew M{c}Gregor]{Andrew M{c}Gregor$^\dag$} 
\thanks{$^\dag$ University of
Pennsylvania, 3330 Walnut Street, Philadelphia, PA19104. E-mail
{\tt andrewm@cis.upenn.edu}. Research supported in
part by  NSF grants CCR-031096 and ITR 0205456}

\begin{abstract} We address the problem of bounding below the probability
of error under maximum likelihood decoding of a binary code with a
known distance distribution used on a binary symmetric channel.  An
improved upper bound is given for the maximum attainable exponent of
this probability (the reliability function of the channel).
In particular, we prove that the ``random coding exponent'' is the true
value of the channel reliability for codes rate $R$ in some interval
immediately below the critical rate of the channel. An analogous result
is obtained for the Gaussian channel.
\end{abstract}

\maketitle

\section{Introduction}

Optimizing $P_e(C,p)$ over all codes of a given rate $R$ has received
much attention in information and coding theory. It is known that for
the best possible codes this probability declines as an exponential
function of the code length. Let us define the largest attainable
exponent of the error probability
\[
E(R,p)=\limsup_{n\to\infty} \frac{1}{n} \log \max_{C\subseteq
\{0,1\}^n, R(C)=R} \frac 1{P_e (C,p) }
\]
also called the error exponent or the reliability of the channel.  The
problem of bounding the function $E(R,p)$ for the binary symmetric and
other communication channels was one of the central problems of
information theory in its first decades. In particular, the standard
textbooks \cite{bla87,csi81,gal68,vit79} all devote considerable attention
to properties and bounds for channel reliability.  There are a
variety of methods for deriving upper and lower estimates of
$E(R,p)$. The most successful approaches to lower bounds are averaging
over a suitably chosen ensemble of codes (for instance, all binary
codes or all linear codes) \cite{gal68}
and relying on the  distance distribution of an average code in a code
ensemble \cite{gal63}, \cite{pol94a}. Recently the distance
distribution approach  was the subject of several papers because of
the renewed interest to  performance estimates of specific code
families (rather than ensemble average estimates).

The problem of upper bounds on the error exponent $E(R,p)$ also has a
long history.  Several important ideas in this problem were suggested
in the paper \cite{sha67}.  The nature of the upper bounds is
different for low values of $R$ and for $R$ close to capacity. For low
code rates paper \cite{sha67} suggested to bound the error probability
below by the probability of making an error to a closest neighbor of
the transmitted codeword.  \remove{Incremental improvements to the
upper bounds on $E(R,p)$ were related to improved upper bounds on the
minimum distance of the code of a given rate $R$. }

\subsection{Notation and previous results}\label{sec:notation}
Since our main result is a new bound on the error exponent $E(R,p)$,
in this section we overview the known bounds on this function. It
should be noted that the method below applies to the analysis of any
code sequence for which the distance distribution is known or can be
estimated.

For notational convenience we shall write $d_{ij}$ for the Hamming
distance between two codewords $x_i$ and $x_j$. We shall write
$d_{iy}$ for the distance between a code word $x_i$ and an arbitrary
word $y$. Let $B^i_w=|\{x\in C:\, d_{ix}=w\}|$ and let $B_w=\sum_i B^i_w/M$
be the local and average distance distributions of the code $C$ of size $M$.

Let $h(x)$ be the binary entropy and $h^{-1}(x)$ its inverse function.
Denote by $\dgv(R):=h^{-1}(1-R)$ the relative Gilbert-Varshamov
distance corresponding to $R$ and by
\[
D(x\|y)=x\log \frac xy +(1-x)\log \frac {1-x}{1-y}
\]
the information divergence between two binomial distributions (the
base of logarithms is 2 throughout). Let
\begin{equation}\label{eq:Aw}
A(\omega): =\omega\log 2\sqrt{p(1-p)},
\end{equation}
$\varphi(x)=h(\half - \sqrt{x(1-x)}).$ Throughout $w=\omega n$, 
$l=\lambda n$ and $d=\delta n$. Let $[n]=\{1,2,\dots,n\}.$ 

For a given $p$, define
   $$
    \rho=\rho(p)=\frac{\sqrt{p}}{\sqrt{p}+\sqrt{1-p}}.
   $$
The function
\[
E_{\text{sp}}(R,p)=D(\dgv(R)\|p)
\]
is called the {\em sphere packing exponent\/}; it gives an upper bound
on $E(R,p)$ which is valid for all code rates $R\in[0,1-h(p)]$ and
tight for code rates $R\ge \rcrit,$ where the value $\rcrit=1-h(\rho)$ is called the critical rate of the channel.
For low rates the best known results for a long time were given by the
following theorem.
\begin{theorem}\label{thm:ld}
\begin{equation}\label{eq:me}
-A(\dgv(R)) \le E(R,p)\le -A(\bar \delta).
\end{equation}
\end{theorem}
Here the lower bound is Gallager's ``expurgation exponent''
\cite{gal63} obtained for instance for a sequence of linear codes
whose minimum distance meets the Gilbert-Varshamov bound. The upper
bound in (\ref{eq:me}) is due to \cite{mce77b}. It is obtained by
substituting the result  of \cite{mce77a} into the ``minimum-distance
bound'' of \cite{sha67}.  The function  $\bar\delta=\dlp(R)$ is the
linear programming bound of \cite{mce77a} on the relative distance of codes of
rate $R$ defined as
\[
\bar\delta :=\min_{\substack {0\le\alpha\le \frac12}}
G(\alpha,\tau)
\]
where $G(\alpha,\tau)=2\frac{\alpha(1-\alpha)-\tau(1-\tau)}
{1+2\sqrt{\tau(1-\tau)}},$ and where $\tau$ satisfies
$h(\tau)=h(\alpha)-1+R.$ Note that Theorem \ref{thm:ld} implies that
$E(0,p)=-A(1/2)$.

Let 
  $$
    \tau_\nu(\xi):=\frac12\Big( 1-
        \sqrt{1- 4\big(\sqrt{\nu(1-\nu)-\xi(1-\xi)}-\xi\big)^2}\Big).
  $$
Let $\bar R(\delta)$ be the inverse function of $\bar\delta(R)$,
  $$
    \bar R(\delta)=1+\min_{(\half)(1-\sqrt{1-2\delta})\le \alpha\le \half}
          (h(\tau_\alpha(\delta/2))-h(\alpha)).
  $$

Derivation of improved upper bounds on $E(R,p)$ 
is based on the following inequality 
for the error
probability $P_e(x_i)$ conditioned on transmission of  the codeword
$x_i$. For every $j\ne i$ let
  $$X_{ij}\subset \{y\in X: d_{jy}\le d_{iy}\}$$ 
be an arbitrary
subset. Let $C'\subset C$ be an arbitrary subcode of $C$ such that
$x_i\not\in C'.$  Then
 \begin{equation}\label{eq:kou} P_e(x_i)\ge \sum_{x_j\in
C'} \Big\{\probi {X_{ij}}-\!\!\!\sum_{x_k\in
C'\backslash\{x_j\}}\!\!\! \probi {X_{ij}\cap X_{ik}}\!\Big\}.
\end{equation} 
 Let us take $C'$ to be the set of codeword neighbors of
$x_i$ at distance $w$ from it. We have, for any $w$,
    $$
      P_e(x_i)\ge B_w^i\probi{X_{ij}}\Big[1- (B_w-1)\probi {X_{ik} |
   X_{ij}}\Big]_+\Big\}, 
    $$
where $x_j, x_k$ are any codewords 
such this $d_{ij}=d_{kj}=w, d_{jk}=d,$ where $d$ is the code's 
minimum distance, and $[a]_+= \max(a,0).$ 
Summing both sides of the last inequality on $i$ from $1$ to $M$,
we obtain the estimate
of $P_e(C)$ in the form 
   \begin{equation}\label{eq:kou1} P_e(C)\gtrsim
   \max_w \Big\{B_w\probi {X_{ij}}
     \\\times\Big[1- (B_w-1)\probi {X_{ik} |
   X_{ij}}\Big]_+\Big\}, 
   \end{equation} 

Recall from \cite{sha67} that a straight-line segment that connects a
point on $E_{\text{sp}}(R',p)$ with a point on any other upper bound
on $E(R,p), R<R'$ is also a valid upper bound on $E(R,p).$ This result
is called the {\em straight-line principle}. It is usually applied in
situation when there is a $\cup$-convex upper bound on $E(R,p)$ and
results into the straight-line segment  given by the common tangent to
this bound and the curve $E_{\text{sp}}(R,p).$

\bigskip
{\sc The results of \cite{lit99}.} The upper bound in (\ref{eq:me})
was  improved in \cite{lit99} by relying on estimates of the distance
distribution of the code.  
The proof in \cite{lit99} is composed of two steps. 
The first part is bounding the distance distribution of
codes by a new application of the linear programming method (similar
ideas were independently developed in \cite{ash99a}). The second step
is using (\ref{eq:kou}) to derive a bound on the error exponent.
The estimate of the distance distribution of codes of \cite{lit99} 
has the following form.
\begin{theorem}{\rm \cite{lit99}} \label{thm:mu} For any family of codes of 
sufficiently large length and rate $R,$ any $\alpha\in [0,1/2]$ and
any $\tau$ that satisfies $0\le h(\tau)\le h(\alpha)-1+R,$ 
there exists a value $\omega, 0\le\omega\le G(\alpha,\tau)$ such that
$n^{-1}\log B_{\omega n}\ge \mu (R,\alpha,\omega)-o(1),$ where 
\begin{equation}
\mu (R,\alpha,\omega)= R-1+h(\tau)+2h(\alpha)-2q(\alpha,\tau,\omega/2)
\\-\omega-(1-\omega) h\Big(\frac{\alpha-\omega/2}{1-\omega}\Big),
\end{equation}
and where
\begin{equation}\label{eq:ii}
q(\alpha,\tau,\omega)=h(\tau)+\int_0^\omega dy
\log\frac{P+\sqrt{P^2-4Qy^2}}{2Q},
\end{equation}
where $P=\alpha(1-\alpha)-\tau(1-\tau)-y(1-2y), Q=(\alpha-y)(1-\alpha-y),$
is the exponent of the Hahn polynomial $H^{\alpha n}_{\tau n}(\omega
n).$
\end{theorem}

The bound on $E(R,p)$ in \cite{lit99} 
has the following form.
\begin{theorem}\label{thm:tl}
\begin{equation}\label{eq:bl}
E(R,p)\le \min_{\alpha,\tau}\;\max_{0\le \delta\le \bar\delta
}\; \max_{\delta\le\omega\le G(\alpha,\tau)} \;N
\end{equation}
\medskip where
\begin{equation}\label{eq:N}
N=\min\{-A(\delta),-\min(\mu(R,\alpha,\omega),\\-B(\omega,\delta))-
         A(\omega)\},
\end{equation}
$0\le \tau\le h^{-1}(h(\alpha)-1+R), 0\le\alpha\le 1/2;$
$A(w)$ is defined in {\rm (\ref{eq:Aw})},
\begin{align}\label{eq:Bwl}
B(\omega,\lambda) &= -\omega-(1-\omega)h(p)+  \max_{\eta\in
[\frac{\lambda p}{2},\min(\frac{\lambda}{4},p(1-\omega))]} \Big (
\lambda \ent{\frac{2\eta}{\lambda}}+(\omega-\lambda/2)
\ent{\frac{\omega-2\eta}{2\omega-\lambda}}\\ &+
(1-\omega-\lambda/2)\ent{\frac{p(1-\omega)-\eta}{1-\omega-\lambda/2}}
\Big ).
\end{align}
\end{theorem}
{\em Remark.}  In \cite{lit99}, optimization in (\ref{eq:bl}) involves
taking a maximum on $\alpha$ and $\tau$. 
However, Theorem \ref{thm:mu} is valid for any 
$\alpha\in[0,1/2],\tau\in[0,h^{-1}(h(\alpha)-1+R)],$ and therefore, 
a better bound is generally obtained by taking
a minimum rather than a maximum. Throughout the rest of the paper
we will assume that $h(\tau)=h(\alpha)-1+R$. 
This assumption simplifies the analysis somewhat and does not seem
to affect the final answer. 

\bigskip
Analysis of the inequality (\ref{eq:kou1}) together with some
additional ideas gives rise to Theorem \ref{thm:tl} and its
improvements.  We begin with deriving a simplified form of the bound
(\ref{eq:bl}) for low rates $R$.

\subsection{A study of the bound (\ref{eq:bl})}
 By omitting the term $A(\delta)$ in (\ref{eq:N}), 
the expression for $N$ can be written as
   $$
    N=\max\{-\mu(R,\alpha,\omega)-A(\omega),B(\omega,\delta)-A(\omega)\}.
   $$
As will be seen below, for low rates $R$, the first term under the
maximum is the greater one. For this reason we begin with the study
of the first term for low rates. Since this term does not depend on $\delta,$
we have
   \[
    \max_{0\le \delta\le \bar\delta
}\; \max_{\delta\le\omega\le G(\alpha,\tau)}(-\mu-A(\omega))
\le \max_{0\le\omega\le G(\alpha,\tau)}(-\mu-A(\omega))
  \]

\begin{lemma}\label{lemma:lp1} Let $p\ge 0.037, 0\le R\le \varphi(\delta_1),$
where $\delta_1=2\rho(1-\rho).$ Then 
 \begin{equation}\label{eq:union-a}
    \max_{0\le\omega\le G(\alpha,\tau)}(-\mu-A(\omega))= -A(\bar\delta )-R+1-h(\bar\delta ).
  \end{equation}
\end{lemma}
\begin{proof}
In the expression $-\mu(R,\alpha,\omega)-A(\omega)$
let us take $\alpha$
equal to the value that furnishes the minimum in the definition of
$\bar\delta.$ Under the assumptions of the lemma,
$R\le 0.303.$  In this case, it is known that $\alpha=1/2$ and
the expression $q(\alpha,\tau,\omega/2)$ simplifies as follows. The
integral in (\ref{eq:ii}) upon a substitution $\alpha=\frac12, 2y=z$
takes the form
  \begin{align*}
   \int_0^{\omega/2}&
\log\frac{P+\sqrt{P^2-4Qy^2}}{2Q}dy\\&=
     \frac12\int_0^\omega\!\! \log 
           \Big[\frac{(1-2\tau)^2
     +\sqrt{(1-2\tau)^2((1-2\tau)^2-4z(1-z))}
                   -2z(1-z)}{2(1-z)^2}\Big]dz\\
     &=\int_0^\omega \log \frac{1-2\tau+\sqrt{(1-2\tau)^2-4z(1-z)}}
                    {2(1-z)}dz.
   \end{align*}              
Let
   $$
    k(\tau,\omega)=
          h(\tau)\\
   +\int_0^\omega \log \frac{1-2\tau+\sqrt{(1-2\tau)^2-4z(1-z)}}
                    {2(1-z)}dz.
  $$
It is known \cite{kal95} that in the region 
$0\le \omega \le(1/2)-\sqrt{\tau(1-\tau)}$, this function gives the
exponent of the Krawtchouk polynomial $K_{\tau n}(\omega n)$, i.e., 
  $$
    \log K_{\tau n}(\omega n)=  n(k(\tau,\omega)+o(1)).
  $$
Therefore, we
obtain the identity $q(1/2,\tau,\omega/2)=k(\tau,\omega)$.
Substituting this in $\mu$ we obtain the following 
  $$
  - \mu-A(\omega)=-2h(\tau)+2k(\tau,\omega)-\omega \log \sqrt{4p(1-p)}.
  $$
Let $g(\omega)=\frac{\partial}{\partial\omega}(-\mu-A(\omega))$. 
From the equation $g(\omega)=0$ we find that the maximizing argument 
$\omega$ satisfies
   $$
    1-2\tau-2 \sqrt u(1-\omega)=-\sqrt{(1-2\tau)^2-4\omega(1-\omega)},
   $$
where $u=\sqrt{4p(1-p)}$. This equation has a real zero if
   $$
     \omega\le \bar \omega:= 1-\frac{1-2\tau}{2\sqrt u},
   $$
and then the maximizing argument is
  $$
    \omega_\ast(\tau)=\frac{\sqrt u}{1+\sqrt u}\Big(1-\frac{2\tau}{1-\sqrt u}
        \Big).
  $$
Recall that $0\le\omega\le G(\half,\tau)=\frac 12-\sqrt{\tau(1-\tau)}.$
We shall show that 
   \begin{equation}\label{eq:disc}
       \arg\max_{0\le\omega\le G(\half,\tau)} (-\mu-A(\omega))=G(1/2,\tau).
   \end{equation}
There are two cases.

$(i).$ Let $R=\phi(\delta_1).$ In this case the stationary 
point $\omega_\ast$ is exactly at the right end of the interval, 
i.e., $\omega_\ast(\tau)=\frac 12-\sqrt{\tau(1-\tau)}.$ To show this, 
compute
  $$
    \delta_1=2\rho(1-\rho)=\frac u{1+u}
  $$
  $$
      \tau=h^{-1}(R)=\frac 12-\sqrt{\delta_1(1-\delta_1)}
               =\frac{(1-\sqrt{u})^2}{2(1+u)},
  $$
and substituting this into $\omega_\ast,\bar\omega$ we find
  $$
    \omega_\ast(\tau)=\frac{\sqrt u}{1+\sqrt u}\Big(
          1-\frac{1-\sqrt u}{1+{u}}\Big)=\delta_1=\bar\omega.
  $$

$(ii).$ Now consider code rates 
$0\le R< \varphi(\delta_1).$ Observe that $\tau=h^{-1}(R)$ decreases
as $R$ decreases, and therefore $\bar\omega$ also decreases with $R$.
On the other hand $\omega_\ast(\tau)$ increases as $\tau$ falls,
so in this case $\bar\omega<\omega_\ast,$ and $g(\omega)$ has no 
zeros for $0\le\omega\le G(\half,\tau).$ It is positive throughout because
$g(0)>0.$ This again proves (\ref{eq:disc}).

Hence, $-\mu-A(\omega)$ increases on $\omega$ for all $\omega\in[0,G],$
attaining the maximum at the right end of this segment. Substituting
$\omega=G(\half,\tau)$ into this expression, we obtain the claim of the lemma.
\end{proof}

For $R \ge 0.305$ the minimum in the definition of $\bar\delta $ is
given by some $\alpha<1/2.$ Fixing $\alpha$ equal to this value we
observe that the function $\mu$ depends only on $\omega.$ Therefore,
the behavior of the function $-\mu(R,\alpha,\omega)-A(\omega)$
can be studied numerically (for instance, using Mathematica).
We observe that this function increases on $\omega$ for $\omega\le \bar\delta(R)$ as long as $R\le \bar R(\delta_1)$. For $R=\bar R(\delta_1),$
the maximum of $-\mu(R,\alpha,\omega)-A(\omega)$ on $\omega$ is attained
for $\omega=\bar \delta=\delta_1.$ Substituting $\omega=\bar\delta $
into $\mu$, we again arrive at the expression (\ref{eq:union-a}).

To summarize, the bound ({\ref{eq:bl}) implies the following: let 
$R\le \bar R(\delta_1),$ then
\begin{equation}\label{eq:2}
    E(R,p)\le \max \Big\{-A(\bar\delta)-R+1-h(\bar\delta),
               \\ \max_{\delta,\omega}(-B(\omega,\delta)-A(\omega))\Big\}.
\end{equation}
Next we show that for low code rates the maximum in this expression is 
given by the
term $-A(\bar\delta)-R+1-h(\bar\delta)$. This is difficult to verify
analytically because of the complicated form of the term $B$; however
this can be verified numerically for any given value of the
probability $p$. More precisely, there exists a value of the rate $R=R_0$,
a function of $p$, 
such that for $0\le R\le R_0$, the first term is 
(\ref{eq:2}) is greater than the second one. 
   
As a result, we obtain the following proposition.
\begin{proposition}\label{prop:sim}
Let $\bar R(\delta_1)\le R_0.$ Then
\begin{equation}\label{eq:union-a1}
E(R,p)\le -A(\bar\delta )-R+1-h(\bar\delta )  \quad 0\le R\le R_0
\end{equation}
\begin{equation}\label{eq:union-b}
E(R,p)\le \max\limits_{0\le \delta\le \bar\delta }
\max\limits_{\delta\le\omega\le \bar\delta }  (
B(\omega,\delta)-A(\omega) )  \quad R_0\le R.
\end{equation}
\end{proposition}

The example of $p=0.01$ is shown in Fig. \ref{fig:errbnds}.

Some comments are in order.  The first term on the right in
(\ref{eq:kou}) is the ``reverse union bound'' which suggests to
estimate the error rate $P_e(x_i)$ by a sum of pairwise error
probabilities. An interesting fact is that for large $n$ and for
certain values of $R$ and $p$ the union bound argument gives the
correct value of the error exponent. From  (\ref{eq:union-a1}) we can
see this and more, namely that for large $n$ and code rates below
$R_0$, the error exponent is given by the sum of pairwise
probabilities of incorrect decoding {\em to a codeword at the minimum
distance of the code $C$} from the transmitted codeword. (Note that
the relative minimum distance of $C$ is bounded above by $\bar\delta
$.) The improvement of  (\ref{eq:union-a1}) over the upper bound in
(\ref{eq:me}) is in that it takes into account decoding errors to all
$\exp(n(R-1+h(\bar\delta )))$ neighbors of the transmitted vector as
opposed to just one such neighbor in (\ref{eq:me}).  The main question
addressed below is to determine the range of code rates where the
union bound and (\ref{eq:union-a1}) is true and to refine the
inequality (\ref{eq:kou}) for those rates where the union bound does
not apply.

In general terms, the answer to this question for large $n$ is given by 
(\ref{eq:kou1}). The bound 
$P_e(C)\gtrsim B_{w}\probi{X_{ij}}$ is valid as long as 
\begin{equation}\label{eq:un1}
     B_{w}\probi{X_{ik}\cap X_{ij}}\lesssim \probi{X_{ij}}.
\end{equation}

In our analysis we use the estimation method of
\cite{bur84}-\cite{bur00} which was originally developed for codes on
the sphere in $\reals^n$. Below we modify it for use in the Hamming
space and improve the estimate (\ref{eq:bl}).  The analysis of the
relation between the distance distribution and $P_e(C,p)$ for the
Hamming space turns out to be more difficult than for $\reals^n$. One
of the issues to be addressed is the choice of decision regions in the
estimation process. We suggest one choice which while still being
tractable leads to improving the estimates.

The results of the present paper are twofold: first, we  expand the
applicability limits of the bound (\ref{eq:union-a1}).  Outside these
limits we will derive a bound on $E(R,p)$ which is better than the
result obtained from Theorem \ref{thm:tl}.

\section{A New Bound}
\subsection{Statement of the result} 
Let us state a lower bound for the error probability of max-likelihood decoding
of an {\em arbitrary sequence of codes} with a given distance
distribution. 

\begin{theorem}\label{thm:new} Let $(C_i)_{i\ge 1}$ be a sequence of codes
with rate $R$, relative distance $\delta$ 
and distance distribution satisfying $B_{\omega n}\ge
2^{n\beta(\omega)-o(n)},$ where $\beta(\omega)>0$ for all 
$\delta\le\omega\le 1.$ 
The error probability of max-likelihood
decoding of these codes satisfies $P_e(C,p)\ge 2^{-En-o(n)},$ where
\begin{equation}\label{eq:new}
E= \min_{\delta\le\omega\le 1}\; \max_{\delta\le \lambda\le \omega}\;
\big[\max (-\beta (\omega) -A(\omega) ,  B(\omega,\lambda)
-A(\lambda))\big]
\end{equation}
where $A$ and $B$ are defined as in Equations {\rm(\ref{eq:Aw})} and
{\rm (\ref{eq:Bwl})} respectively.
\end{theorem}

Theorem \ref{thm:new} will be proved later in this section.
We first discuss its application to the problem of bounding 
$E(R,p)$.  Let us specify
this theorem for the distance distribution defined by Theorem \ref{thm:mu}.
Let $\alpha,\tau, G(\alpha,\tau)$ have the same meaning as in
(\ref{eq:bl}). Recall that by Theorem \ref{thm:mu} for 
any family of codes of rate
$R$ and every $\alpha\in[0,1/2]$ there exists an $\omega,
0\le\omega\le G(\alpha,\tau)$ such that the average number of neighbors
at distance $\omega n$ can be bounded as 
$B_{\omega n}\ge 2^{n\mu(R,\alpha,\omega)-o(n)}.$ Let us substitute this
distance distribution in (\ref{eq:new}) and perform optimization.
By Lemma \ref{lemma:lp1} and the argument after it, for low values of
$R$ we conclude that the function $E(R,p)$ is bounded above by 
(\ref{eq:union-a}). Let $R_0^\ast$ be the value of the rate,
a function of $p$, for which
the maximum shifts from the first term in (\ref{eq:new}) to
the second one. As in the previous section, we arrive at the following
theorem.

\begin{theorem}\label{thm:nb} Let $\bar R(\delta_1)\le R_0^\ast.$
Then
\begin{equation}\label{eq:new-a}
E(R,p)\le -A(\bar\delta )-R+1-h(\bar\delta ) \quad 0\le R\le R_0^\ast
\end{equation}
\begin{equation}\label{eq:new-b}
E(R,p)\le\max\limits_{0\le \lambda \le \bar\delta }
\max\limits_{\lambda \le\omega\le \bar\delta }
B(\omega,\lambda)-A(\lambda)  \quad R_0^\ast \le R,
\end{equation}
where $A$ and $B$ are defined as in Equations {\rm(\ref{eq:Aw})} and
{\rm (\ref{eq:Bwl})} respectively.
\end{theorem}
\remove{\begin{proof} (outline)

Hence in (\ref{eq:new}) we need to compute the maximum on $\omega.$
For the same reason we must maximize on $\lambda \in (0,G(\alpha,\tau)).$ 
Hence the bound (\ref{eq:new}) in this case takes on the form
  $$
E(R,p)\le \min_{0\le\alpha\le 1/2}\;\max_{0\le \lambda\le G(\alpha,\tau)}\;
\max_{\lambda\le\omega\le G(\alpha,\tau)} \;
\big[\max (-\mu (R,\alpha,\omega) -A(\omega) , 
B(\omega,\lambda) -A(\lambda))\big].
  $$
The result is now obtained by performing an analysis similar to that of 
the previous section.
\end{proof}}

\medskip
{\bf Example.} (Explanation of Fig.~\ref{fig:errbnds}) 
To show that (\ref{eq:new}) improves over
(\ref{eq:bl}), let $p=0.01.$ Then from
(\ref{eq:union-a1})-(\ref{eq:union-b}) we obtain $R_0\approx 0.271.$
 From (\ref{eq:new}) we find that the bound  (\ref{eq:union-a1}) is
valid for $R\le R_0^\ast\approx 0.388.$ Note also that 
$\rcrit=0.559, \bar R(\delta_1)=0.537.$ See Figure~\ref{fig:errbnds}
for a graph of the known error bounds including our new bounds.  In
the figure,  curve (a) is a combination of the best  lower bounds on
the error exponent. Curve (b) is the union bound of
(\ref{eq:union-a1}), (\ref{eq:new-a}).  Curve (c) is the upper bound
(\ref{eq:union-b}) given by Theorem \ref{thm:tl},
Prop. \ref{prop:sim}.  Curve (d) is the upper bound (\ref{eq:new-b})
given by Theorem \ref{thm:new}. Curve (e) is the sphere-packing bound
$E_{\text{sp}}(R,p)$.

The improvement of Theorem \ref{thm:new} over Theorem \ref{thm:tl} is
in the extended region where the union bound (a) is applicable and in
a better bound for greater values of the rate $R$.

Note that $E_{\text{sp}}(R,p)$  is better than (b) from $R\approx
0.422$; the straight-line bound  (not shown) further improves  the
results.

Another set of examples together with some implications of Theorems
\ref{thm:new}-\ref{thm:nb} will be given in Sect.~\ref{sec:straight}.

\medskip
{\bf Remark.} Experience leads us to believe that the maximums in the
equation are achieved for $\omega=\lambda = \bar\delta $ which would
give  us the bound
\begin{equation*}
E(R,p)\le \left \{
\begin{array}{ll}
-A(\bar\delta )-R+1-h(\bar\delta ) & 0\le R\le R_0^\ast\\ B(\bar\delta
,\bar\delta )-A(\bar\delta )  & R_0^\ast \le R.
\end{array}
\right .
\end{equation*}
However this has proved too difficult to verify analytically due to
the cubic condition for $\eta$ in the maximization term in the definition
of $B(\omega,\lambda)$ and other computational problems.

\begin{figure}[tH]
\epsfxsize=10cm \setlength{\unitlength}{1cm}
\begin{center}
\begin{picture}(6,6)
\put(-1,0){\epsffile{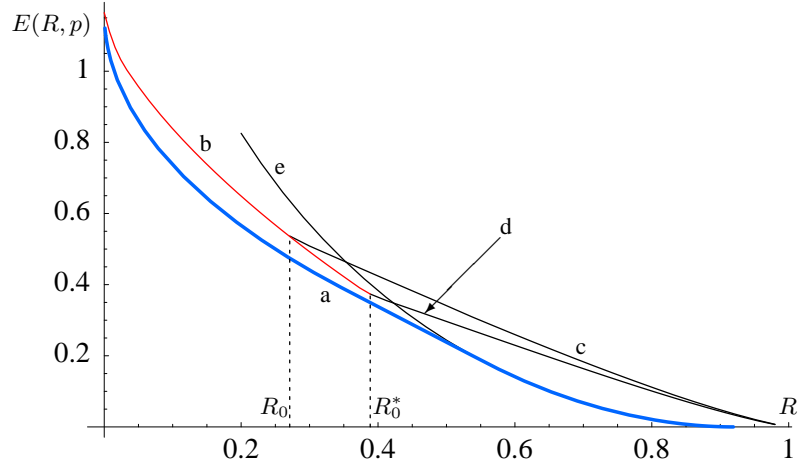}} \put(2.6,2.2){{\footnotesize
a}} \put(1,4.2){{\footnotesize b}} \put(6,1.5){{\footnotesize c}}
\put(5,3.05){\vector(-1,-1){1}} \put(5,3.1){{\footnotesize d}}
\put(2,3.9){{\footnotesize e}}
\put(8.7,0.7){{\footnotesize\mbox{$R$}}}
\put(-1.5,5.8){{\footnotesize\mbox{$E(R,p)$}}}
\put(1.8,0.7){{\footnotesize\mbox{$R_0$}}}
\put(3.3,0.7){{\footnotesize\mbox{$R_0^\ast$}}}
\end{picture}\end{center}
\caption{Bounds on the error exponent for the BSC with
$p=0.01$. Notation explained in the text.}\label{fig:errbnds}
\end{figure} 

\subsection
{Preview of the proof} The basic idea of the estimation method is from
\cite{bur00} although we make some modifications due to the fact that
the observation space is discrete.  To prove this theorem we start by
choosing a collection of sets $\{Y_{ij}\}$, each corresponding to a
pair of codewords $(x_i,x_j)$, such that $Y_{ij}$ is outside the
decoding region of $x_i$ and
\[ Y_{ij}\cap Y_{ik}= \emptyset \mbox{ for all }k\neq j.
\]
Then we can bound the error probability in terms of these sets using
the following inequality
\[
P_e \geq \frac{1}{M} \sum_{i=1}^{M} \sum_{j:d_{ij}=w} \probi{Y_{ij}}
\qquad(w=1,2,\dots,n).
\]
One of the main questions in applying this inequality and further
ideas of \cite{bur00} is the choice of the sets $Y_{ij}$. We construct
the $Y_{ij}$'s via sets $X_{ij} \subset \binf^n,$ where
\[
X_{ij}=\{y \in F^n : d_{iy}=d_{jy}= \frac{d_{ij}}{2}+p(n-d_{ij})\}.
\]

\begin{figure}[tH]
\begin{center}
\epsfig{file=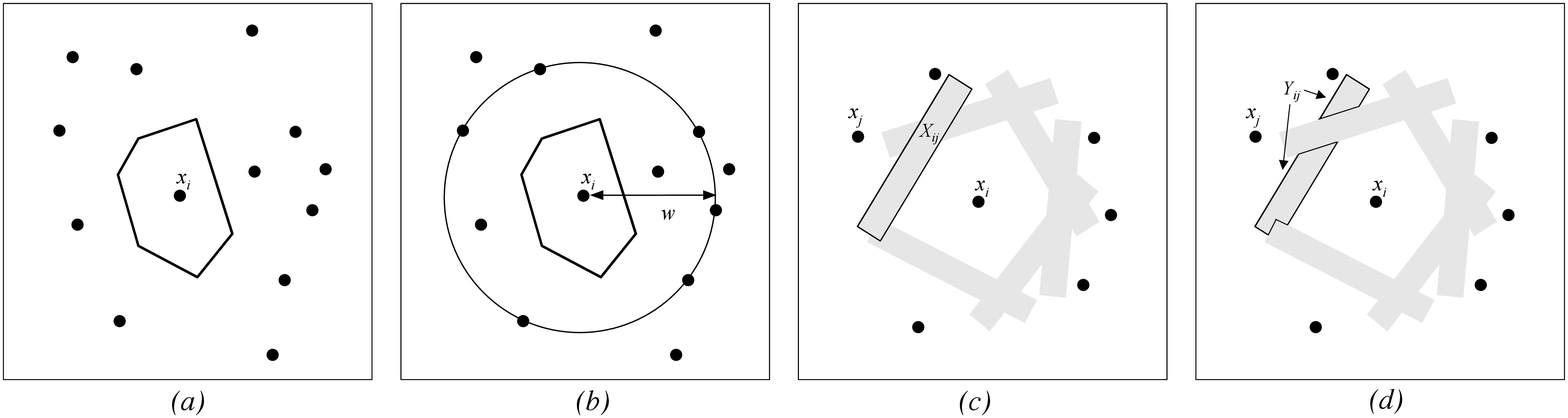,width=6.4in,height=1.6in}
\caption{The Bounding Process. (a) A codeword $x_i$, neighboring
codewords and the Voronoi region $D(x_i)$. (b) We restrict our
attention to only those neighbors that are a distance $w$ away. By
only worrying that the received word $y$ is closer to this subset of
the neighbors we upper bound  $\probi{D(x_i)}$. (c) For each neighbor
$x_j$ still under consideration,  let $X_{ij}$ be some set of words
that are closer to $x_j$ than they are to $x_i$. (d) We ``prune'' the
$X_{ij}$'s to construct disjoint $Y_{ij}$'s with the required
properties.  }
\label{fig:bnddiag}
\end{center}
\end{figure} 

See Figure~\ref{fig:bnddiag} for an illustration of the bounding
process.  To create the  $Y_{ij}$'s from the $X_{ij}$'s we randomly
``prune'' these sets so that the disjointness condition is
satisfied. To accomplish this pruning we define a set of codewords
$T_i=\{x_j: d_{ij}=w\}$ for each codeword $x_i$.  
Then, as in \cite{bur00}, for each $x_i$, we randomly index by $s_{ij}$ 
all the codewords $x_j$ that are a distance $w$ from $x_i$. 
Define sets
\[T(i,j)=\{k\in T_i : s_{ik}<s_{ij}\}.\]
We then get our $Y_{ij}$'s as follows
\[
Y_{ij} = X_{ij} \setminus [ \cup_{k\in T(i,j)} X_{ik}].
\]
These $Y_{ij}$ satisfy the disjointness condition: assume there exists
$x\in Y_{im}\cap Y_{il}$. Then $x\in X_{im}$ and $x\not \in
\bigcup_{k\in T(i,m)} X_{ik}$ gives that $s_{il}>s_{im}$. However we
also have $x\in X_{il}$ and $x\not \in \bigcup_{k\in T(i,l)} X_{ik}$
and this gives that $s_{im}>s_{il}$ which is a contradiction.

Instead of calculating  $\probi{Y_{ij}}$ directly we apply a ``reverse
union bound'' to get
\begin{equation}\label{eq:reverse-union}
\probi{Y_{ij}} \geq \probi{X_{ij}}(1-K_{ij}),
\end{equation}
where $K_{ij} =\sum_{k\in T(i,j)} \probi{X_{ik} | X_{ij}}$. Note that
this inequality is the bound (\ref{eq:kou}) with our
particular choice of $X_{ij},Y_{ij}.$  Using the last inequality
we perform a recursive procedure which shows the existence of a subcode
$C'\subset C$ with large error probability (among the codewords of $C'$).
This gives the claimed lower bound on $P_e(C,p).$

\subsection{A proof of Theorem \ref{thm:new}}
The error probability for two codewords is given by the following
well-known lemma.
\begin{lemma}
For all codewords $x_i$ and $x_j$ that are a distance $w$ apart
$\lim_{n\to\infty}  \frac{1}{n} \log  P_i(X_{ij})=A(\omega),$ where
$A(\omega)$ is defined in {\rm (\ref{eq:Aw})}.
\end{lemma}

\begin{lemma}
For all codewords $x_i$, $x_j$ and $x_k$ such that $d_{ij}=d_{ik}=w$
and $d_{jk}=l$ we have 
  $$\lim_{n\to\infty}  \frac{1}{n} \log
\probi{X_{ik}|X_{ij}}=B(\omega,\lambda)
   $$ 
where $B(\omega,\lambda)$ is
defined in Eq.  (\ref{eq:Bwl}).
\end{lemma}
\begin{proof}
First consider
\begin{align*}
\probi{X_{ik} \cap X_{ij}}=&\sum_{m=0}^{ \min(l/2,p(n-w))}
\ncr{l/2}{m}^2 \times \ncr{w-l/2}{w/2-m} \ncr{n-w-l/2}{p(n-w)-m} \\
&\times p^{w/2+p(n-w)}(1-p)^{n-w/2-p(n-w)}.
\end{align*}
Then since
\[\log \probi{X_{ik}|X_{ij}} = 
\log \probi{X_{ik} \cap X_{ij}}-\log \probi{X_{ij}},\]  substituting
for $\probi{X_{ij}}$ from the previous lemma and taking the
appropriate limits gives the required result.
\end{proof}

The following properties of $B(\omega,\lambda)$ can be verified
numerically.

\begin{lemma}
If $\omega\leq \lambda \leq 2\omega$ then $B(\omega,\lambda)\leq
B(\omega,\omega)$.  If $\lambda \leq \omega$ then
$B(\lambda,\lambda)\leq B(\omega,\lambda)$
\end{lemma}

Recall that the indexing of pairs to create the sets $T(i,j)$ is done
randomly. By linearity of expectation there exists an indexing such
that
\begin{equation}
P_e\geq \frac{1}{M} \sum_{i=1}^{M} \sum_{j: d_{ij}=w}
\expec{\probi{Y_{ij}}} \label{eq:expec}
\end{equation}
This equation will be the basis for our new bound on the error
exponent but before deriving this bound we have two final
preliminaries. Firstly we will refer to all codewords $x_j$ that are a distance $w$ from $x_i$ as $w$-neighbors of $x_i$. (Recall that we defined $B_w^i$ to be the number of codewords in the $w$-neighborhood of $x_i$.)  Secondly we shall say that a subset $S'\subseteq S$ of codewords is of
{\em substantial} size (with respect to $S$) if its size has the same exponential order as
the size of $S$. Note that for a family of codes $(C_i)_{i\geq 1}$  where $C_i$
has length $n$ and rate $R$, we can consider  $(C_i^{'} )_{i\geq 1}$,
a family of codes where $C'_i$ is a  substantially sized subcode of
$C_i$, when trying to bound the error  exponent since
\[\lim_{n\to\infty} R(C'_i)=\lim_{n\to\infty} R(C_i)=R\]
and
\[\limsup_{n\to\infty} \frac{1}{n} \log \frac{1}{P_e (C_i^{'},p)} \geq
\limsup_{n\to\infty} \frac{1}{n} \log \frac{1}{P_e (C_i,p)}. \]

We now proceed with a case analysis dependent on the values of $K_{ij}$. Roughly speaking when $K_{ij}$ is typically less than a half, a union bound argument will be used to bound the error  probability. When $K_{ij}$ is typically larger than a half, a more complicated analysis will be required. Before we describe the two cases in our analysis we need the following two lemmas.

\begin{lemma} \cite{bur01a} Suppose that there are $L$ balls of 
$K$ different colors. The number of balls of a color $k$ is
$r_k$. We are also given numbers $n_k, 1\le k\le K$. 
Suppose that all balls are enumerated randomly 
by different integers from 1 up to $L$. Let $\tau$ be a random integer
between 1 and $L$ and let
$t_k$ be the number of balls of color $k$ with numbers between 1
and $\tau$. Then
\[
\prob{ t_k \leq n_k, k=1,\ldots K} \geq 
\min\Big\{1,\frac{1}{4} \min_{1\leq k \leq
K}
\frac{n_k}{r_k}\Big\}.
\]
\end{lemma}
Recall that, for a given $(i,j)$ pair, $K_{ij}$ is a random variable.
We then can prove the following lemma:
\begin{lemma}
Let $d_{ij}=\omega n$. With respect to the random indexing of all 
the $(i,k)$ pairs (where $x_k$ is any codeword such that $d_{ik}=\omega n$) 
we have
  \[\prob{K_{ij}\leq \frac{1}{2}} \geq 
       \min\Big\{1,
\min_{l\in \Lambda}
    \frac{2^{-nB(\omega,\lambda)-o(n)}}{\min\{B_w^i,B_l^j\}}\Big\}\]
where $\Lambda = \{l \in [n] : |R_{w,l}|>N_{w,l}\}$, $R_{w,l}=\{x_k\in C
: d_{ij}=d_{ik}=w, d_{jk}=l\} $ and  $N_{w,l} =
\frac{2^{-nB(\omega,\lambda)}}{2(n+1)}$.
\label{lem:pij}
\end{lemma}

\medskip\begin{proof}
\begin{align*}
\mathbb{P}(K_{ij} &\leq 1/2)
=
\prob{
\sum_{k\in T(i,j)} \probi{X_{ik} |X_{ij}} \leq 1/2}
\\&\cong
\prob{
\sum_{l=0}^{n}
\sum_{k\in T(i,j), d_{jk}=l} 2^{nB(\omega,\lambda)}   \leq 1/2}\\
&= 
\prob{
\sum_{l=0}^{n} |T(i,j)\cap R_{w,l}|  2^{nB(\omega,\lambda)} \leq 1/2}\\[2mm]
&\geq 
\prob{
|T(i,j)\cap R_{w,l}|  \leq N_{w,l}\; \forall l\in \Lambda}.
\end{align*}
Let there be a ball for each codeword in $\bigcup_l R_{w,l}.$ Consider a
ball from $R_{w,l}$ to have color $l$. Let $n_l=N_{w,l}$
and $ \mu_l=|x_m\in R_{w,l}: s_{im}<s_{ij}|.$ We have
\[\prob{K_{ij}\leq 1/2} \geq \prob{\mu_l\leq n_l \;\forall l \in \Lambda
}.\]
By the previous lemma we have
\[\prob{\mu_l\leq n_l \;\forall l \in \Lambda } \geq \frac{1}{4}
\min_{l\in \Lambda} \frac{n_l}{|R_{w,l}|}
\]
if the right-hand side is less than one.
The lemma then follows from the fact that $|R_{w,l}|\leq \min\{B_w^i, B_l^j\}$.
\end{proof}

\medskip

In the analysis that leads to Theorem 6, we face a dichotomy of a 
relatively sparse $w$-neighborhood of the transmitted 
vector $x_i$ when the union bound is asymptotically tight, 
and a cluttered neighborhood when is not. These two cases
correspond to the first and the second terms in (\ref{eq:new}), respectively.
When the union bound analysis is not applicable, we will rely crucially 
on the following lemma.

\begin{lemma}\label{gammaij}
If $K_{ij}>1/2$ for some $i,j$ such that $d_{ij}=\omega n$ then there 
exists a nonempty set $\Lambda_{ij}$ such that for all  
$\lambda\in \Lambda_{ij}$, 
    $$  
  \min\{B_w^i,B_{\lambda n}^j\}> 2^{-nB(\omega,\lambda)-o(n)}.
   $$
\end{lemma}

\medskip\begin{proof}
Consider a pair of codewords $x_i$ and $x_j$ such that  $K_{ij}> 1/2$.  
We deduce
that $\prob{K_{ij}\leq 1/2}<1$ since the event $\{K_{ij}> 1/2\}$ occurred.
Therefore,
by Lemma \ref{lem:pij}, there exists a $\lambda$ such that,
\[
    \frac{2^{-nB(\omega,\lambda)-o(n)}}{\min\{B_w^i,B_{\lambda n}^j\}}<1.\]
\end{proof}

Given a pair of codewords $x_i,x_j$ with $K_{ij}\leq 1/2$ we put $\Lambda_{ij}
=\emptyset;$ otherwise, we assume that $\Lambda_{ij}$ contains all the
values of $\lambda=l/n$ whose existence is established in the
previous lemma. We now
define, for all $n$ possible values of $l=\lambda n$, the sets
   $$
    G_{l,w}=\{x_j: \exists x_i \mbox{ such that } K_{ij}>\half
        \text{ and }l/n\in \Lambda_{ij}\}.
   $$
In words, for a given $l$, the set $G_{l,w}\subset C$ contains all the
codewords $x_j$ that have a $w$-neighbor $x_i$ 
such that the set $\Lambda_{ij}$ contains the value $\lambda=l/n.$
Let $H_{l,w}$ be defined as the set of all $x_i\in C$
such that a  substantial number of the $w$-neighbors $x_j$ of $x_i$ 
satisfy $K_{ij}>\half$ and $l/n \in\Lambda_{ij}.$
Note that the ``substantial number'' here is in relation to $B^i_w$.

We say $\lambda=l/n$ is a ``nuisance level" for $\omega$ if $H_{l,w}$ and
$G_{l,w}$ are both substantially sized subcodes of $C$. The two cases in
the following analysis correspond to whether or not a nuisance level
exists. The next theorem bounds the error probability in the case that
it does not exist.

\begin{theorem}\label{thm:firpart}
Consider any code $C$ of sufficiently large length $n$ and rate $R$.
Assume that for some $\omega$ and bounding function $f$ we have
$\frac{1}{n} \log B_{\omega n}^i \geq f(\omega) $ for all $i$.
If there does not exist a nuisance level for $\omega$ then
\[\frac{1}{n} \log \frac{1}{P_e(C,p)} \leq -f(\omega) - A(\omega)+o(1).
\]
\end{theorem}

\medskip\begin{proof}  
Let us define the sets
\[S_1=\{l:H_{l,w} \mbox{ is not a substantially sized sub-code} \},\]
\[S_2=\{l:G_{l,w} \mbox{ is not a substantially sized sub-code} \}.\]
Since $w$ does not have a nuisance level, $S_1\cup S_2=[n]$.
Without loss of generality we may assume that $G_{l,w}=\emptyset$ for
all $l\in S_2$ since removing $\bigcup_{l\in S_2} G_{l,w}$ yields a
substantially sized subcode. Hence also $H_{l,w}=\emptyset$ for all
$l\in S_2$.  Now consider only transmitting the codewords in
$C'=C\setminus \bigcup_{l\in [n]} H_{l,w}$ and note that this is a
substantially sized number of codewords since  neither
$\bigcup_{l\in S_1} H_{l,w}$ nor $\bigcup_{l\in S_2} H_{l,w}$ are 
substantially sized. For each of these codewords we know that
$\frac{1}{n} \log B_{\omega n}^i \geq f(\omega) $. Hence
\begin{align*}
       P_e(C,p) &\geq \frac{1}{M}\sum_{i=1}^{M} \sum_{j:d_{ij}=w}
   \probi{Y_{ij}}\\ 
      &\gtrsim \frac{1}{M}\sum_{x_i\in C'} B_w^i \min_{j:d_{ij}=w}
   \{\probi{Y_{ij}}\}\\ 
            &\geq \frac{ 1}{2}  \min_{i,j:d_{ij}=w} (B_w^i
    \probi{X_{ij}})\\ & \geq 2^{n(A(\omega)+f(\omega) )-o(n)}.
\end{align*}
The second inequality follows from the fact that for each $x_i\in C'$, 
a substantial number of $w$-neighbors $x_j$ are such that $K_{ij}\leq 1/2$,
and the third one is implied by (\ref{eq:reverse-union}) since
$\probi{Y_{ij}}\geq \probi{X_{ij}}/2$ whenever $K_{ij}\leq 1/2$.
\end{proof}

We now bound the error probability (and ensure another property of 
the distance distribution) in the case that there exists a nuisance level.

\begin{theorem} \label{thm:secpart}
Consider any code $C$ of sufficiently large length $n$ and rate $R$
and an $\omega\in [0,1]$.   
Let $\lambda$ be a nuisance level for $\omega.$ 
The subset of codewords $x_j\in C$ such that
  $$
     |\{x_k\in C: \dist(x_j,x_k)=\lambda n\}|\ge 2^{-n
B(\omega,\lambda)-o(n)}$$
forms a substantially sized subcode. 
Furthermore,
\[
\frac{1}{n} \log \frac{1}{P_e(C,p)} \leq  B(\omega,\lambda)- A(\omega)+o(1).
\]
\end{theorem}

\medskip\begin{proof}
Since $G_{l,w}$ is substantially sized, it follows by
Lemma~\ref{gammaij} that a substantial number of codewords $x_j$ have
at least $2^{-nB(\omega,\lambda)-o(n)}$ neighbors at a relative
distance $\lambda$.  Now consider $x_i\in H_{l,w}.$
By definition, there is a substantially sized
subset $N(i)$ of the $\omega n$-neighbors of $x_i$
such that $\lambda \in
\Lambda_{ij}$ for all  $x_j\in N(i)$.  Hence, appealing to
Lemma~\ref{lem:pij}, for each $x_j\in N(i)$,
\[\prob{K_{ij}\leq 1/2} \geq
\frac{2^{-nB(\omega,\lambda)-o(n)}}
{B_w^i}.
\]

Now
\begin{align*}
\expec{\probi{Y_{ij}}} &=   \expec{I_{K_{ij}\leq \frac{1}{2}}
\probi{Y_{ij}}}
+\expec{I_{K_{ij}> \frac{1}{2}} \probi{Y_{ij}}}\\
&\geq
\expec{I_{K_{ij} \leq \frac{1}{2}} \probi{Y_{ij}}}\\
&\geq
\frac{2^{nA(\omega)}}{2}\prob{K_{ij}\leq \frac{1}{2}},
\end{align*}
and so, by the above discussion and Eq. (\ref{eq:expec}), we get
\begin{align*}
P_e &\geq  \frac{1}{M}\sum_{x_i\in H_{l,w}} \sum_{j\in N(i)}
\expec{\probi{Y_{ij}}}\\
&\geq 
\frac{1}{M}\sum_{x_i\in H_{l,w}}
2^{nA(\omega)} 2^{-nB(\omega,\lambda)-o(n)}
\\
&= 2^{n(A(\omega) -
B(\omega,\lambda))-o(n)}.
\end{align*}
\end{proof}

\begin{proof}[Proof of Theorem \ref{thm:new}.]
Let $C$ be the code from the statement of the theorem. Let
  $$
    F=\frac{1}{n} \log \frac{1}{P_e(C,p)}.
  $$
As discussed in \cite{ash00c}, \cite{bur00}, for any $w=\omega n,
\delta\le\omega\le 1,$ the code 
$C$ contains a subcode $C'$ of size $M'\ge M/n^2$
such that for all codewords $x_i$ in this subcode
\[
\frac{1}{n}\log B_{\omega n }^i >\beta(\omega)-o(n).
\]
Since the subcode is substantially sized we may now consider this 
subcode as our new code.

For a fixed $\omega$ construct $Y_{ij}, X_{ij}$ and $K_{ij}$ for 
all $(i,j)$ pairs with $d_{ij}=\omega n$. By Theorems 
\ref{thm:firpart} and \ref{thm:secpart} we get
\[
F  \leq \left \{
\begin{array}{l@{\qquad}l}
-\beta(\omega)- A(\omega)&\mbox{ if no nuisance level exists for $\omega$}\\
B(\omega,\lambda_1)- A(\omega)&\mbox{ if a nuisance level $\lambda_1$ exists for $\omega$.}\\
\end{array}
\right .
\]
Hence we get
\[
F
\leq 
\max \{-\beta(\omega), B(\omega,\lambda_1)\}- A(\omega).
\]

Now if $\lambda_1\geq \omega$ then $B(\omega,\lambda_1)\leq
B(\omega,\omega)$ and so we get
\begin{equation} \label{eq:nonuis}
F
\leq 
\max \{-\beta(\omega), B(\omega,\omega)\}- A(\omega).
\end{equation}
If $\lambda_1 < \omega$ then we use the fact from Theorem \ref{thm:secpart} 
that for a substantial number of codewords $x_i$, 
$B_{\lambda_1n}^i \geq 2^{-nB(\omega,\lambda_1)}$. We now construct 
new $Y_{ij}, X_{ij}$ and $K_{ij}$ for all $(i,j)$ pairs with 
$d_{ij}=\lambda_1 n$. Hence by 
Theorems \ref{thm:firpart} and \ref{thm:secpart} we get
\[
F  \leq \left \{
\begin{array}{l@{\qquad}l}
B(\omega,\lambda_1)- A(\lambda_1)
& \mbox{ if no nuisance level exists for $\lambda_1$}\\
B(\lambda_1,\lambda_2)- A(\lambda_1)
&\mbox{ if a nuisance level $\lambda_2$ exists for $\lambda_1$.}
\end{array}
\right .
\]

Hence we get
\[
F
\leq 
\max \{B(\omega,\lambda_1),B(\lambda_1,\lambda_2)\}- A(\lambda_1).
\]

If $\lambda_2\geq \lambda_1$ then $B(\lambda_1,\lambda_2)\leq
B(\lambda_1,\lambda_1)\leq B(\omega,\lambda_1)$ then
\[
F
\leq B(\omega,\lambda_1)- A(\lambda_1).
\]
If $\lambda_2< \lambda_1$ then we use the fact that for a substantial number of codewords $x_i$, $B_{\lambda_2 n }^i
\geq
2^{-nB(\lambda_1,\lambda_2)}$ and continue as before.

We continue in this manner and get a sequence
$\omega>\lambda_1>\lambda_2\ldots$ such that at step $i$ we get the bound
\[
F
\leq \max \{B(\lambda_{i-1},\lambda_i),
B(\lambda_i,\lambda_{i+1})\}- A(\lambda_i).
\]
This process terminates after at most $n$ steps since there are only $n$
possible values for the nuisance level. At the last step, $i=f$,
the nuisance level $\lambda_{f +1}$, if it even exists, is not less than
$\lambda_{f}$ itself and therefore we have
\begin{align*}
F &\leq \max \{B(\lambda_{f -1},
\lambda_{f }), B(\lambda_{f },
\lambda_{f +1})\}- A(\lambda_f)\\
&\leq \max \{B(\lambda_{f-1},
\lambda_{f }),
B(\lambda_{f },\lambda_{f })\} - A(\lambda_f) \nonumber\\
&\leq  B(\omega,\lambda_{f})- A(\lambda_f). \label{eq:salt}
\end{align*}
Now for our code either this equation or Eq. (\ref{eq:nonuis}) is valid,
and so we have shown that for every $\omega,\delta\le\omega\le 1$ 
there exists  
$\lambda\leq \omega$ such that 
\[
F
\leq  \max(-\beta(\omega)-A(\omega), B(\omega,\lambda)- A(\lambda)).\]
This completes the proof.
\end{proof}

\section{More on the bound of Theorem (\ref{thm:nb})}\label{sec:straight}

In this section we take a closer look at the bound (\ref{eq:new-a})
with the aim to show that it provides a new segment of code rates where
the BSC channel reliability is known exactly. We rely on the notation
of Sect.~\ref{sec:notation}. Let $R_x=1-h(2\rho(1-\rho)).$
Recall that the best known
lower bound on $E(R,p)$ below the critical rate is given by 
  \begin{equation}
      E_x(R,p)=-A(\dgv(R)) \quad 0\le R\le R_x
  \end{equation}
  \begin{equation}\label{eq:E0}
      E_0(R,p)=D(\rho\|p)+\rcrit-R \quad R_x<R\le \rcrit.
  \end{equation}
For $R>\rcrit$ the reliability function $E(R,p)=E_{\text{sp}}(R,p).$
Note that both $E_x$ and $E_{\text{sp}}(R,p)$ can be viewed as instances
of the union bound and that both are tangent on $E_0(R,p).$ 
Let us make one simple observation showing that the bound (\ref{eq:new-a})
has the same property. 

The following lemma is verified by direct calculation.
\begin{lemma}\label{lemma:tangent}
 Let $\delta_1=2\rho(1-\rho)$ and let $R_1=\bar R(\delta_1).$ 
Then
   $$
     -A(\delta_1)-R_1+1-h(\delta_1)=E_0(R_1,p).
   $$
\end{lemma}
\begin{proof} Indeed, (\ref{eq:E0}) can be rewritten as
  $$
    E_0(R,p)=1-R+\log(1+2\sqrt{p(1-p)}).
  $$
The equality in the statement is equivalent to the relation
   $$
      h(\delta_1)+\delta_1\log 2\sqrt{p(1-p)}=\log(1+2\sqrt
             {p(1-p)})
   $$
which is an easily verifiable identity.
\end{proof}

Next we can prove the main result of this section.
\begin{theorem}\label{thm:tight} Let $p, 0.046 \le p<1/2$ be the channel transition 
probability. Then the channel reliability $E(R,p)$ equals the
random coding exponent $E_0(R,p)$ for $R_1\le R\le \rcrit.$
\end{theorem}
\begin{proof} We check numerically that $R_1<R_0^\ast$ for $p\ge 0.046.$
Thus, by Theorem \ref{thm:nb} 
for these values of $p$ we have $E(R_1,p)=E_0(R_1,p).$
The full claim follows from the straight-line bound of Shannon,
Gallager, and Berlekamp \cite{sha67}.
\end{proof}
{\em Remark.} We have seen in Lemma~\ref{lemma:lp1} that for $p\ge 0.037,$
it suffices to rely on the simple form of the function $\bar R(x),$
namely $R(x)=\varphi(x)$. Thus the only numerical calculation involved in
the proof of this theorem relates to the function $B(\omega,\delta).$

The random coding exponent $E_0(R,p)$ gives the best known lower bound
on $E(R,p)$ for $R_x\le R\le \rcrit.$ The fraction of this segment
in which Theorem \ref{thm:tight} shows it to be tight is given by
   $$
     \frac {\rcrit-R_1}{\rcrit-R_x}.
   $$
This fraction equals about $1/3$ for $p=0.05$ and tends to one as 
$p\to \half.$

We give an example of the new picture for the $E(R,p)$ function
in Fig.~\ref{fig:rel}.
Previously the reliability of the BSC 
was known exactly only for $R\ge \rcrit$ \cite{eli55a}.

\begin{figure}[tH]
\vspace*{2cm}
\epsfxsize=10cm \setlength{\unitlength}{1cm}
\begin{center}
\begin{picture}(6,6)
\put(-1,0){\epsffile{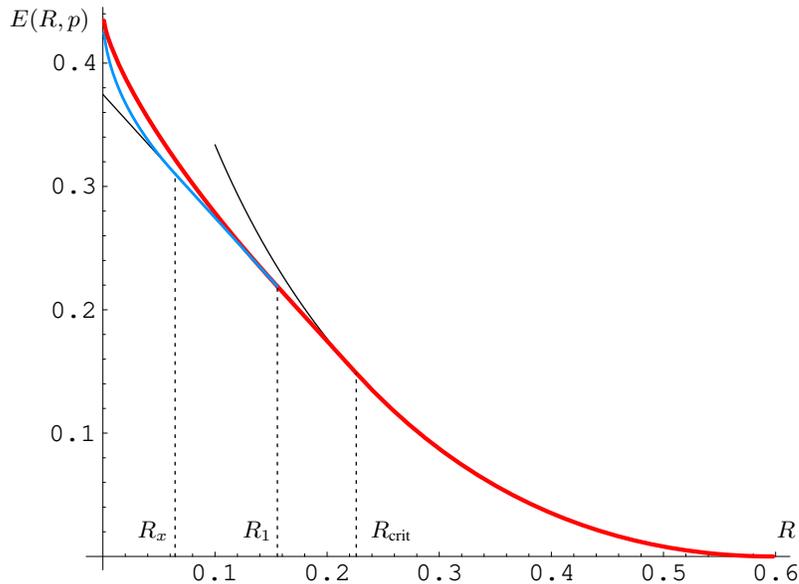}} 
\put(8.7,0.7){{\footnotesize\mbox{$R$}}}
\put(0.2,0.7){{\footnotesize\mbox{$R_x$}}}
\put(-1.5,7.5){{\footnotesize\mbox{$E(R,p)$}}}
\put(1.6,0.7){{\footnotesize\mbox{$R_1$}}}
\put(3.3,0.7){{\footnotesize\mbox{$\rcrit$}}}
\end{picture}\end{center}
\caption{Bounds on the error exponent for the BSC with
$p=0.08$. In the interval $R_1\le R\le \rcrit$ the random
coding bound $E_0(R,p)$ is tight.}\label{fig:rel}
\end{figure} 

\section{Random linear codes}
The inequality of Theorem \ref{thm:new} can be used for a code with
an arbitrary distance distribution. In this section
we are interested in the estimate of the error exponent for a random
linear code $C$. Here by a random code we mean a binary code whose 
weight distribution behaves as the binomial distribution:
${\mathcal B}_{\omega n}\cong \exp[n(R+h(\omega)-1)].$
The reason for calling this code random is that the weight distribution
of a randomly chosen linear code with high probability
converges to the binomial distribution (e.g. \cite{bar02b}).

The error exponent $\tilde E(R,p)$
for random linear codes for low rates is bounded
below by the expurgation exponent: $\tilde E(R,p)\ge -A(\dgv(R)).$ 
For $R_x\le R_0\le\rcrit,$ the exponent $\tilde E(R,p)\ge E_0(R,p).$
Moreover, it is known that the error probability $P_e(C)$ averaged
over the ensemble of all binary
codes meets this bound with equality \cite{gal73}.
The proof of this result in \cite{gal73} is accomplished by
computing the ensemble average probability of error under
list decoding into lists of size 2, where by error we mean
the event that the transmitted codeword is not in the resulting list.
It turns out that under this definition the error occurs in an 
exponentially smaller fraction of cases than the error of maximum likelihood
decoding. In other words, in all the cases of error under maximum likelihood
decoding (i.e., decoding into a size-1 list) except for an exponentially
small fraction of them, there is exactly one codeword which is
at least as close to the received word as is the transmitted word.
This shows that for exponential asymptotics of the error probability
of random codes the union bound is tight. An analogous result can also
be proved for the ensemble of binary linear codes.

Here we compute a lower bound on the decoding error probability of a
code with weight distribution ${\mathcal B}_{\omega n}.$ 
A closed-form expression
again seems beyond reach, however computational evidence with the
bound (\ref{eq:new}) suggests that in a certain segment of code
rates $0\le R\le R^{\ast\ast}$, 
the error exponent of maximum likelihood decoding
of the code $C$ is bounded above as follows 
\[
\tilde E(R,p)\le -A(\dgv(R)).
\]
In other words, the expurgation exponent is tight for a random linear code
in the region of low code rates.

\section{The Gaussian channel}
Given the results for the BSC of Section \ref{sec:straight}, 
it is natural
to assume that qualitatively similar results hold for the reliability
function of the Gaussian channel.
Here we consider briefly this problem and show that the random coding exponent
is tight for a certain interval of rates immediately below the
critical rate. As in the binary case, the length of this segment
depends on the level of the channel noise.

Let $a$ be the signal-to-noise ratio in the channel.
Denote by $E(R,a)$ the channel
reliability function defined analogously to the BSC case. It is known
to be bounded below by the random coding bound $E_0(R,a)$ \cite{sha59}
which has the form
  $$
     E_0(R,a)=\frac a4(1-\cos\theta_x)+R_x-R
  $$
and is the best known lower bound for $R_x\le R\le \rcrit$
where
  $$
    R_x=\frac12\ln\Big(\frac12+\frac12\sqrt{1+\frac {a^2} 4}\,\Big),
  $$
  $$  
\theta_x=\cos^{-1}\sqrt{1-e^{-2R_x}},
  $$
  $$
    \rcrit=\frac12\ln\Big(\frac12+\frac a4+\frac12\sqrt{1+\frac {a^2} 4}\,\Big).
  $$
Let $C$ be a code on $S^{n-1}(\reals)$ (the unit sphere in $\reals^n$).
 Let $\theta(x_i,x_j)$
be the angle between the vectors that correspond to the codewords
$x_i,x_j$. Denote by $B(\theta)$ the distribution of angular 
distances in the code $C$. The exponent of the union bound on the error
probability $P_e(C,a)$  has the form
  $$
  E_U=\frac a4 (1-\cos\theta)-\frac1n\ln B(\theta).
  $$
Used together with an estimate of the distance
distribution of a code of rate $R$ obtained in \cite{ash00c}
this bound takes the form
  \begin{equation*}
   E_U(R,a)=\frac a4 (1-\cos\bar\theta)-\ln \sin\bar\theta -R,
  \end{equation*}
where $\bar\theta=\bar\theta(R)$ is the root of the 
equation $R=\psi(\theta)$ and
  $$
   \psi(x)=-\frac{1-\sin x }{2\sin x }\ln\frac{1-\sin x }{1+\sin x }
              -\ln\frac{2\sin x }{1+\sin x}
  $$
(which represents the Kabatiansky-Levenshtein bound on spherical codes).
The strongest known condition for the union bound to be valid asymptotically 
as a lower bound on  $P_e(C,a)$ was announced in \cite{bur01b}.
According to it, $E(R,a)\le E_U(R,a)$ for all rates $R\le R^\ast$,
where $R^\ast$ is the root of
  \begin{equation}\label{eq:cond}
    R+\ln\sin\bar\theta(R)=\frac a8(1-\cos\bar\theta).
  \end{equation}
Other conditions were obtained in \cite{ash00c,bur00,coh04}. 

Next we state a result analogous to Lemma \ref{lemma:tangent}. Its proof
is immediate by comparing the expressions for $E_U$ and $E_0.$
\begin{lemma} Let $R_1=\psi(\theta_x),$ then $E_0(R_1,a)=E_U(R_1,a).$
\end{lemma}
We conclude that $E_0(R_1,a)$ is the correct value of 
$E(R_1,a)$ if $R_1\le R^\ast.$
The last inequality holds for $0<a\le 5.7.$ Coupled with the straight-line
principle of \cite{sha67} this gives 
\begin{theorem} Let $0<a\le 5.7$ be the signal-to-noise ratio in the channel. 
Then 
  $$E(R,a)=E_0(R,a) \qquad (R_1\le R\le R_c).$$
\end{theorem}
\noindent{\bf Example.} For instance, let $a=2$. Then $R_x=0.094, R_1=0.199,
R^\ast=0.263, \rcrit=0.267$.

If instead of (\ref{eq:cond}) we rely on conditions with a published proof,
we would still be able to make a tightness claim of $E_0$ but for a smaller
segment of the signal-to-noise ratio values.

\bigskip
{\em Postscriptum\/}: Recently, a generalized de Caen inequality was 
used to derive lower estimates of error probability
of a code via its distance distribution 
\cite{coh04}. In particular, \cite{coh04} gives a condition for the
union bound to be valid asymptotically as a lower bound on $P_e$ 
in the BSC case.
Although the condition is stated as an optimization problem 
(\cite{coh04}, Prop.~5.3), computational evidence suggests that its 
solution is given by (\ref{eq:un1}). Thus, the methods of this paper and
of \cite{coh04}, although different in nature, seem to lead to the same
general estimates. Note that \cite{coh04} does not contain results on the
BSC reliability function.

\renewcommand\baselinestretch{0.9}
{\footnotesize\bibliographystyle{amsplain}

\begin{thebibliography}{10}

\bibitem{ash99a}
A.~Ashikhmin and A.~Barg, \emph{Binomial moments of the distance distribution:
  {B}ounds and applications}, IEEE Trans. Inform. Theory \textbf{45} (1999),
  no.~2, 438--452.

\bibitem{ash00c}
A.~Ashikhmin, A.~Barg, and S.~Litsyn, \emph{A new upper bound on the
  reliability function of the {G}aussian channel}, IEEE Trans. Inform. Theory
  \textbf{46} (2000), no.~6, 1945--1961.

\bibitem{bar02b}
A.~Barg and G.~D. Forney, Jr., \emph{Random codes: {M}inimum distances and
  error exponents}, IEEE Trans. Inform. Theory \textbf{48} (2002), no.~9,
  2568--2573.

\bibitem{bla87}
R.~E. Blahut, \emph{Principles and practice of information theory},
  Addison-Wesley, Reading, MA, 1987.

\bibitem{bur01b}
M.~V. Burnashev, \emph{On relation between code geometry and decoding error
  probability}, Proc. 2001 IEEE Internat. Sympos. Inform. Theory, Washington,
  DC, p.133.

\bibitem{bur84}
\bysame, \emph{A new lower bound for the $\alpha$-mean error of parameter
  transmission over the white {G}aussian channel}, IEEE Trans. Inform. Theory
  \textbf{30} (1984), no.~1, 23--34.

\bibitem{bur00}
\bysame, \emph{On the relation between the code spectrum and the decoding error
  probability}, Problems of Information Transmission \textbf{36} (2000), no.~4,
  3--24.

\bibitem{bur01a}
M.~V. Burnashev and Y.~A. Kutoyants, \emph{On minimal $\alpha$-mean error
  parameter transmission over a {P}oisson channel}, IEEE Trans. Inform. Theory
  \textbf{47} (2001), no.~6, 2505--2515.

\bibitem{coh04}
A.~Cohen and N.~Merhav, \emph{Lower bounds on the error probability of block
  codes based on improvements of de Caen's inequality}, IEEE Trans. Inform.
  Theory (2004), no.~2, 290--310.

\bibitem{csi81}
I.~Csisz{\'a}r and J.~K{\"o}rner, \emph{Information theory. {C}oding theorems
  for discrete memoryless channels}, Akad{\'e}miai Kiad{\'o}, Budapest, 1981.

\bibitem{cae97}
D.~de~Caen, \emph{A lower bound on the probability of a union}, Discrete Math.
  \textbf{169} (1997), no.~1-3, 217--220.

\bibitem{eli55a}
P.~Elias, \emph{Coding for noisy channels}, IRE Conv. Rec., Mar. 1955,
  pp.~37--46. Reprinted in D. Slepian, Ed., Key papers in the development of
  information theory, IEEE Press, 1974, pp. 102--111.

\bibitem{gal63}
R.~G. Gallager, \emph{Low-density parity-check codes}, MIT Press, Cambridge,
  MA, 1963.

\bibitem{gal68}
\bysame, \emph{Information theory and reliable communication}, John Wiley \&
  Sons, New York e.a., 1968.

\bibitem{gal73}
\bysame, \emph{The random coding bound is tight for the average code}, IEEE
  Trans. Inform. Theory (1973), no.~2, 244--246.

\bibitem{kal95}
G.~Kalai and N.~Linial, \emph{On the distance distribution of codes},
IEEE Trans. Inform. Theory \textbf{41} (1995), no.~5, pp.~1467-1472. 

\bibitem{ker00}
O.~Keren and S.~Litsyn, \emph{A lower bound on the probability of error on a
  bsc channel}, The 21st IEEE Convention of the Electrical and Electronic
  Engineers in Israel, 2000, pp.~217--220.

\bibitem{kou68}
E.~G. Kounias, \emph{Bounds for the probability of a union, with applications},
  Ann. Math. Statist. \textbf{39} (1968), 2154--2158.

\bibitem{kua00a}
H.~Kuai, F.~Alajaji, and G.~Takahara, \emph{A lower bound on the probability of
  a finite union of events}, Discrete Math. \textbf{215} (2000), no.~1-3,
  147--158.

\bibitem{kua00b}
\bysame, \emph{Tight error bounds for nonuniform signalling over {AWGN}
  channels}, IEEE Trans. Inform. Theory \textbf{46} (2000), no.~7, 2712--2718.

\bibitem{lit99}
S.~Litsyn, \emph{New upper bounds on error exponents}, IEEE Trans. Inform.
  Theory \textbf{45} (1999), no.~2, 385--398.

\bibitem{mce77b}
R.~J. McEliece and J.~K. Omura, \emph{An improved upper bound on the block
  coding error exponent for binary-input discrete memoryless channels}, IEEE
  Trans. Inform. Theory \textbf{23} (1977), no.~5, 611--613.

\bibitem{mce77a}
R.~J. McEliece, E.~R. Rodemich, H.~Rumsey, and L.~R. Welch, \emph{New upper
  bound on the rate of a code via the {D}elsarte-{M}ac{W}illiams inequalities},
  IEEE Trans. Inform. Theory \textbf{23} (1977), no.~2, 157--166.

\bibitem{pol94a}
G.~Sh. Poltyrev, \emph{Bounds on the decoding error probability of binary
  linear codes via their spectra}, IEEE Trans. Inform. Theory \textbf{40}
  (1994), no.~4, 1284--1292.

\bibitem{seg98}
G.~E. S{\'e}guin, \emph{A lower bound on the error probability for signals in
  white {G}aussian noise}, IEEE Trans. Inform. Theory \textbf{44} (1998),
  no.~7, 3168--3175.

\bibitem{sha59}
C.~E. Shannon, \emph{Probability of error for optimal codes in a {G}aussian
  channel}, Bell Syst. Techn. Journ. \textbf{38} (1959), no.~3, 611--656.

\bibitem{sha67}
C.~E. Shannon, R.~G. Gallager, and E.~R. Berlekamp, \emph{Lower bounds to error
  probability for codes on discrete memoryless channels, {II}}, Information and
  Control \textbf{10} (1967), 522--552.

\bibitem{vit79}
A.~J. Viterbi and J.~K. Omura, \emph{Principles of digital communication and
  coding}, McGraw-Hill, 1979.

\end{thebibliography}

\providecommand{\bysame}{\leavevmode\hbox to3em{\hrulefill}\thinspace}

}

\end{document}